\documentclass[useAMS,usenatbib,usegraphicx]{mnras}

\def\aj{AJ}%
\def\araa{ARA\&A}%
\def\apj{ApJ}%
\def\apjl{ApJ}%
\def\apjs{ApJS}%
\def\ao{Appl.~Opt.}%
\def\aap{A\&A}%
\def\mnras{MNRAS}%
\def\prd{Phys.~Rev.~D}%
\def\pasj{PASJ}%
\def\nat{Nature}%
\usepackage{graphicx}
\usepackage[latin1]{inputenc}
\usepackage{color}
\usepackage{times}
\usepackage{natbib}
\usepackage{setspace}
\usepackage{hyperref}
\newif\ifAMStwofonts
\AMStwofontstrue
\definecolor{red}{rgb}{1,0.,0.}

\def\lesssim{\lower.5ex\hbox{$\; \buildrel < \over \sim \;$}}
\def\gtrsim{\lower.5ex\hbox{$\; \buildrel > \over \sim \;$}}
\title[High-z QLF evolution] {Eddington accreting Black Holes in the
  Epoch of Reionization.} \author[Fontanot et al.]{
  \parbox[t]{\textwidth}{Fabio Fontanot$^{1,2}$\thanks{E-mail:
      fabio.fontanot@inaf.it}, Stefano Cristiani$^{1,2,3}$, Andrea
    Grazian$^4$, Francesco Haardt$^{5,6}$, Valentina
    D'Odorico$^{1,2,7}$, Konstantina Boutsia$^8$, Giorgio
    Calderone$^1$, Guido Cupani$^1$, Francesco Guarneri$^{9,1}$,
    Chiara Fiorin$^{10}$, Giulia Rodighiero$^{10,4}$}
    \vspace*{8pt}\\
    $^1$ INAF - Astronomical Observatory of Trieste, via G.B. Tiepolo 11, I-34143 Trieste, Italy \\
    $^2$ IFPU - Institute for Fundamental Physics of the Universe, via Beirut 2, 34151, Trieste, Italy \\
    $^3$ INFN - National Institute for Nuclear Physics, via Valerio 2, I-34127, Trieste, Italy \\
    $^4$ INAF - Osservatorio Astronomico di Padova, Vicolo dell'Osservatorio 5, I-35122, Padova, Italy \\
    $^5$ DiSAT, Universit\`a dell'Insubria, Via Valleggio 11, I-22100 Como, Italy\\
    $^6$ INFN, Sezione di Milano-Bicocca, Piazza delle Scienze 3, I-20123 Milano, Italy \\
    $^7$ Scuola Normale Superiore, Piazza dei Cavalieri, I-56126 Pisa, Italy \\
    $^8$ Las Campanas Observatory, Carnegie Observatories, Colina El Pino, Casilla 601, La Serena, Chile \\
    $^9$ Dipartimento di Fisica, Sezione di Astronomia, Universit\`a  di Trieste, via G.B. Tiepolo 11, I-34131 Trieste, Italy \\
    $^{10}$ Dipartimento di Fisica e Astronomia, Universit\`a di Padova, Vicolo dell'Osservatorio, I-3 35122 Padova, Italy \\
}

\begin{document}
\date{Accepted ... Received ...}

\maketitle

\begin{abstract} 
The evolution of the luminosity function (LF) of Active Galactic
Nuclei (AGNs) at $z \gtrsim 5$ represents a key constraint to
understand their contribution to the ionizing photon budget necessary
to trigger the last phase transition in the Universe, i.e. the epoch
of Reionization. Recent searches for bright high-z AGNs suggest that
the space densities of this population at $z>4$ has to be revised
upwards, and sparks new questions about their evolutionary paths. Gas
accretion is the key physical mechanism to understand both the
distribution of luminous sources and the growth of central
Super-Massive Black Holes (SMBHs). In this work, we model the high-z
AGN-LF assuming that high-z luminous AGN shine at their Eddington
limit: we derive the expected evolution as a function of the
``duty-cycle'' ($f_{\rm dc}$), i.e. the fraction of life-time that a
given SMBH spends accreting at the Eddington rate. Our results show
that intermediate values ($f_{\rm dc} \simeq 0.1$) predict the best
agreement with the ionizing background and photoionization rate, but
do not provide enough ionizing photons to account for the observed
evolution of the hydrogen neutral fraction. Smaller values ($f_{\rm
  dc} \lesssim 0.05$) are required for AGNs to be the dominant
population responsible for Hydrogen reionization in the Early
Universe. We then show that this low-$f_{\rm dc}$ evolution can be
reconciled with the current constraints on Helium reionization,
although it implies a relatively large number of inactive SMBHs at
$z\gtrsim5$, in tension with SMBH growth models based on heavy
seeding.
\end{abstract}


\section{Introduction}\label{sec:intro}                     

The phase transition in the early Universe called {\it Epoch of
  Reionization} (EoR) marks an epoch of major transformation in
baryonic properties, with the large majority of its hydrogen content
moving from a neutral to an ionized state at $z \gtrsim 5.3$ \citep[see
  e.g.][and references herein]{Bosman21}. EoR represents the epoch
when the first complex astrophysical structures, i.e. galaxies, start
to assemble, producing large amounts of stars in the process. As
galaxies grow, Super-Massive Black Holes (SMBH) lying at their very
centre also experience gas accretion, giving rise to the first Active
Galactic Nuclei (AGN) and Quasar (QSO) phenomena. Both star formation
and AGN activity are critical in the production of the ionizing
photons required to drive the Universe outside the so-called {\it Dark
  Ages}.

Results from cosmological probes such as Planck \citep{Planck_cosmpar}
broadly constrain the redshift span of this transition to lie between
$6<z<10$ with peak activity around $z\simeq7$. The development of the
EoR, its overall duration and topology have been the subject of major
discussion in recent years, as these properties are directly linked to
the astrophysical population responsible for the production of
ionizing photons involved in the process. Generally speaking both Star
Forming Galaxies (SFGs) and AGNs are likely contributors to the
ionizing photon budget \citep{Fontanot12b}, however their relative
contribution is still matter of debate. This is not a secondary issue,
as the nature of the dominant sources of ionizing photons are likely
to affect the evolution of the process itself.

We expect that an EoR dominated by SFGs will start earlier and proceed
for a relatively large redshift range ($6 \lesssim z \lesssim 15$)
\citep[see e.g.][]{Bouwens09}: this is due to the fact that SFGs are
numerous sources, but they produce a limited amount of ionizing
photons per each $M_\odot/$yr of stellar mass formed. On the other
hand, AGNs are a rare population, but they efficiently produce
ionizing photons per each $M_\odot/$yr of gas accreted onto the SMBH
\citep{Telfer02, Stevans14}: this implies that an AGNs-driven scenario
favours a late and short EoR, that tends to be in better agreement
with recent findings of a fast drop of both the mean free path of
ionizing photons \citep{Becker21} and the space density of
Ly$_{\alpha}$ emitters \citep{Morales21} at $z>6$.

The estimate of the redshift evolution of the space density of the AGN
population (i.e. its luminosity function - LF - $\Phi$) at $z >
5$ is thus of paramount relevance in order to estimate their relative
contribution to the EoR. Such a goal is not an easy one as the robust
derivation of completeness levels for different surveys is a complex
task with controversial results, even for samples focusing only on the
brightest-end of the LF \citep[see e.g.][among the others]{Jiang16,
  Yang19}: for example \citet{Schindler19} show that the efficient
selection of high-z QSOs in the SDSS does not correspond to high
completeness levels. New efforts have recently allowed us to improve
our understanding of this statistical estimator. The QUBRICS (QUasars
as BRIght beacons for Cosmology in the Southern Hemisphere) survey
\citep{Calderone19, Boutsia20} is a prime example of a reliable QSOs
candidate sample extracted from the combination of several
observational databases (covering the wavelength range from the
optical to the infrared) using machine learning techniques
\citep{Guarneri21}. Several of these candidates have been
spectroscopically confirmed in the last few years (with a success rate
close to 70 percent): the confirmed candidates have been then used to
provide estimates for the bright-end of the AGN-LF at $z \simeq 3.9$
\citep{Boutsia21}. Moreover, again using QUBRICS data,
\citet{Grazian22} estimate the space density for $M_{1450} \simeq -28.6$
AGNs at $4.5<z<5$, and find that it is consistent with a scenario of
a pure density evolution between $z \simeq 3.9$ and $z\simeq 4.75$
with a parameter\footnote{Our reference pure density evolution
  scenario scales with redshift as $\Phi(z) = \Phi(z=4) 10^{\gamma
    (z-4)}$.}  $\gamma=-0.25$. This $\gamma$ value is smaller
(i.e. the evolution is slower) than the corresponding $\gamma=-0.38$
estimated from the ELQS \citep[Extremely Luminous Quasar
  Survey,][]{Schindler19}, and also from the extrapolation of
lower-redshift results based on multi-wavelength surveys
\citep{Shen20}. The evolution of the AGN/QSO-LF represents a key
aspect for models of the ionizing background, as it critically
controls the total number of ionizing photons produced by accretion
onto SMBHs events. \citet{Giallongo15} first suggested (later
confirmed in \citealt{Giallongo19}) that a relatively high space
density of faint AGNs may account for the total photon budget required
for EoR (if these objects retain the same properties - e.g. spectral
slope and escape fraction distribution - of their brighter
counterparts). While finalising our study, preliminary results for AGN
candidates in the JWST Cosmic Evolution Early Release Science Survey
seem to strengthen the case for a high space density of faint AGNs at
$z \simeq 5$ \citep{Onoue23}. Moreover, comparing AGN-LFs with LFs for the
total (inactive) galaxy population holds critical constraints for
models of AGN feedback and their impact on galaxy evolution \citep[see
  e.g.][and reference herein]{Fontanot20}.

While the search for reliable candidates and their spectroscopic
confirmation is routinely performed by several groups both in the
Northern and Southern skies, the finding of very bright QSOs at the
edge of the EoR poses a number of theoretical challenges. Quasars like
J031343.84-180636.4 \citep[z=7.642]{Wang21}, HSC J124353.93+010038.5
\citep[z=7.07]{Matsuoka19}, ULAS J134208.10+092838.61
\citep[z=7.54]{Banados18} or ULAS J112001.48+064124.3
\citep[z=7.085]{Mortlock11} are all powered by SMBHs with estimated
masses within 10$^8$-10$^9 M_\odot$. The mere existence of
such massive structures when the Universe is approximately 750
Myrs old is usually interpreted as an evidence for very efficient
accretion onto SMBHs at early epochs \citep[see e.g.][]{DiMatteo12}.

Prompted by these considerations, in this work we will explore a
simplified model based on the assumption that all SMBHs at $z>5$ (the
redshift range where estimates for the AGN/QSO has been recently
revised upwards) accrete at their Eddington limit for a fraction of
their lifetime (i.e. the so-called ``duty cycle''). We will then
rescale this assumption into predictions for the evolution of the
AGN/QSO-LF, and, consequently, on predictions for the contribution of
the AGN population to the observed ionizing background. The overall
exercise will give us hints on the number of ionizing photon
associated with the early build-up of the more massive SMBHs that are
available for EoR. Throughout the paper we assume a standard $\Lambda$
cold dark matter concordance cosmological model
(i.e. $\Omega_\Lambda=0.7$, $\Omega_m=0.3$, $H_0=70 \, {\rm
  km/s/Mpc}$) and we refer the absolute magnitude at 1450 {\AA}
($M_{1450}$) to the AB system.

\section{Modelling high-z SMBHs evolution}\label{sec:gaea}
In order to estimate the redshift evolution of the QSO-LF we start
from considering different scenarios for the growth of SMBH powering
these luminous sources. For the purpose of the present work we have
tried to adopt the simplest hypotheses allowing us to explore the
general trends, well aware that some of the conclusions can be
circumvented by more complex schemes. In particular, we assume that
whenever a SMBH accretes material at $z\gtrsim5$ it is doing so at the
Eddington rate. We thus write its mass evolution using the following
equation:

\begin{equation}\label{eq:eddaccr}
M_{\rm SMBH}(t) = M_0 \, e^{\big [\frac{(1-\epsilon)}{\epsilon} \frac{f_{\rm dc}}{t_{\rm edd}} t \big ]}
\end{equation}

\noindent
which includes three free parameters, namely the radiative efficiency
$\epsilon$, the Eddington timescale $t_{\rm edd}=4.5 \times 10^8$ yr
and the fraction of time the SMBH is accreting $f_{\rm dc}$ (i.e. its
``duty cycle''). Eq.~\ref{eq:eddaccr} clearly shows that the growth at
Eddington rate depends on both parameters $\epsilon$ and $f_{\rm
  dc}$. Nonetheless, in this study we prefer to fix $\epsilon=0.1$ as
reference value, in order to maintain a physical understanding of our
conclusions as a function of $f_{\rm dc}$. We briefly discuss the
impact of a different choices for $\epsilon$ in the following
sections: although the exact $f_{\rm dc}$ values quoted in the
discussion may change, our conclusions are robust against reasonable
combinations of parameters.

In this work, we will consider a backward approach: we start from
observed AGN-LFs at the highest redshifts accessible with present-day
surveys and we then try to assess the expected evolution to even
higher redshift. In particular, we use as a benchmark the analytical
form for the $z \simeq 5$ AGN-LF, as proposed by \citet{Grazian22}. In
detail, we adopt a fairly standard double power-law approximation for
the LF ($\phi$):

\begin{displaymath}
  \phi(M) = \frac{\phi_\star}{10^{0.4 (M-M_\star) (\alpha+1)} + 10^{0.4 (M-M_\star) (\beta+1)}}
\end{displaymath}

\noindent
with the following parameters ($\alpha$, $\beta$, $M_\star$,
$Log(\phi_\star)$) = (-1.85, -4.065, -26.50, -7.05). First, we use
this LF definition to estimate the mass function of active SMBH at
$z \simeq 5$ (aBHMF), by assuming that all SMBH powering AGNs at these
redshifts shine at their Eddington limit. This simplified assumption
implies that the shape of the aBHMF is identical to the shape of the
AGN-LF by construction, which is in reasonable agreement with the
results from more comprehensive models of AGN synthesis
(\citealt{MerloniHeinz08} - see e.g. their Fig.~5), at least at the
bright/high-mass end. We use this estimate of the aBHMF as a starting
point to reconstruct the aBHMF at higher redshifts using
Eq.~\ref{eq:eddaccr}. We then use these aBHMFs estimates to assess the
AGN-LF and the total BH mass function (BHMF) evolution at different
redshifts.

The choice of a fixed Eddington accretion rate allows us to treat
accretion, luminosities and SMBH masses as equivalent quantities, and
to easily move from one to another. It is worth stressing that other
options, like super-Eddington or sub-Eddington accretion, are possible
for high-z AGNs (and they would have important degeneracies with both
$\epsilon$ and $f_{\rm dc}$). However, including them in our framework
would introduce additional parameters (e.g the Eddington ratio) and
increase the level of degeneracy in our modelling. Moreover, observed
AGNs are not characterized by a given Eddington ratio, but rather by a
distribution of values. Our assumption corresponds to a scenario where
the average value at $z>5$ is close to unity for a wide range of AGN
luminosity, with a relatively small spread.
\begin{figure}
  \centerline{ \includegraphics[width=9cm]{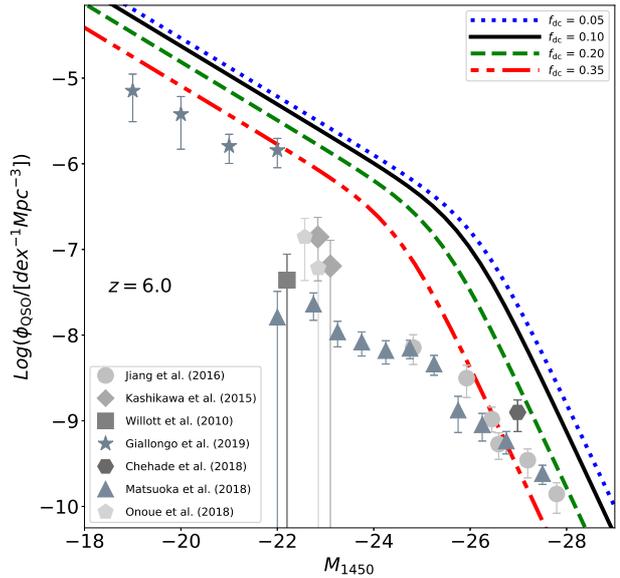} }
  \caption{Estimated $z=6$ AGN LF. The predictions from our models
    are compared with observational constraints. Line types, colours and
    symbols as indicated in the legends.}\label{fig:uplf}
\end{figure}

Larger $f_{\rm dc}$ values imply a faster LF evolution. Vice-versa, a
small duty cycle implies an almost negligible evolution of the space
density of luminous sources. Indeed, in our simplified framework and
for a given SMBH, a short $f_{\rm dc}$ implies a small probability of
being active (and for a short time). Therefore, in order to explain
the observed AGN space densities, we need to assume a large number of
available SMBHs, most of which are expected to be non-accreting. These
``inactive'' SMBHs are available for powering AGN of similar
luminosities at slightly earlier times. On the other hand, a $f_{\rm
  dc} \simeq 1$ implies that the observed $z\simeq5$ AGNs should correspond
(almost) to the only available SMBHs of that mass in the Universe.

It is also worth noticing that, for all $f_{\rm dc}$ values, the
evolution of the AGN-LF strictly follows the Eddington accretion path:
this implies that our estimated evolution qualifies as a {\it pure
  luminosity evolution}, in contrast with the typical pure density
evolution scenario often assumed to estimated the LF evolution
\citep[see e.g.][]{Kim21}. In particular, $f_{\rm dc}>0.2$ scenarios
predict a fast drop in the space density of the brightest
$M_{1450} < -27.5$ QSOs, which results in a negligible number of these
sources at $z\gtrsim7$ (assuming that QSOs at $z=5$ and $z=7$
belong to the same parent population and/or the number of detected
$z\simeq7$ QSOs is representative of the total population). On the other
hand, if $f_{\rm dc}<0.2$ the space density of bright sources evolves
moderately from $z=5$ to $z=7$.

\section{Constraining the model with observations}\label{sec:results}

\subsection{Luminosity Functions}\label{sec:lf}

In order to get a first constraint on our model predictions, in
Fig.~\ref{fig:uplf} we compare them with available estimates for $z
\simeq 6$ AGN/QSO LF \citep{Willott10, Kashikawa15, Jiang16, Onoue17,
  Chehade18, Matsuoka18, Giallongo19}. These data provide us with an
estimate for the space densities of active BHs at different
luminosities. By focusing at the bright end, we may conclude that
$f_{\rm dc} \simeq 0.35$ is a reasonable value that reproduces the
available evidence; at the same time, we can exclude larger $f_{\rm
  dc}$ values that would correspond to a faster than observed $5<z<6$
LF evolution. However, if also $z\simeq6$ data are subject to relevant
incompleteness, as the QUBRICS space densities at $z\simeq5$ suggest, we
can see them as lower limits for the space densities of active BHs,
which translates into a $f_{\rm dc} \lesssim 0.35$. 
Therefore, we conclude that the comparison with available constraints
on space densities favours $0.35 \lesssim f_{\rm dc} \lesssim 0.1$.

\subsection{Ionizing Backgrounds}
\begin{figure*}
  \centerline{ \includegraphics[width=18cm]{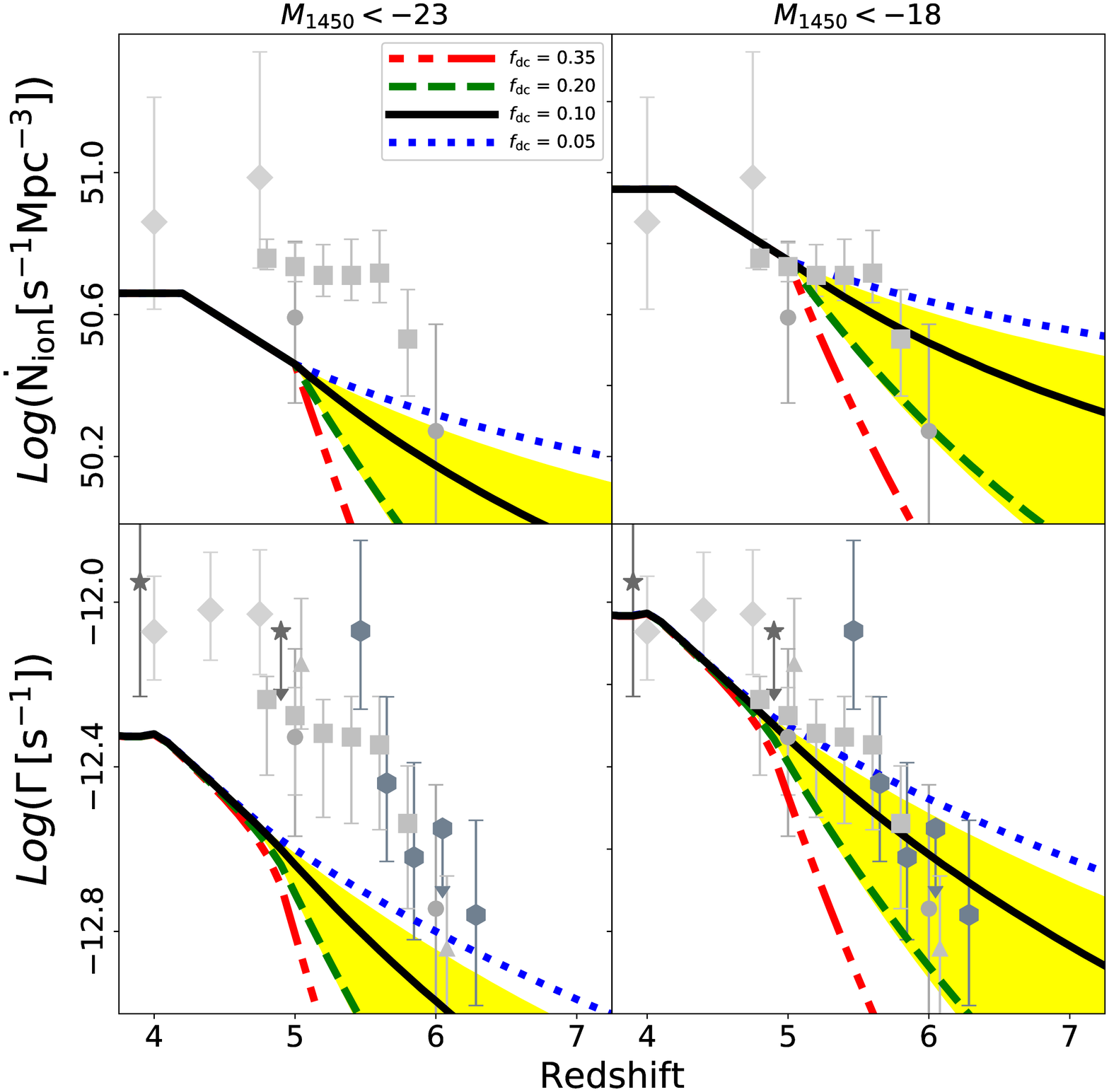} }
  \caption{Redshift evolution of the photon volume emissivity ($N_{\rm
      ion}$ -{\it upper panels}) and hydrogen photoionization rate
    ($\Gamma$ - {\it lower panels}). Observed data are from
    \citet[circles]{WyitheBolton11}, \citet[triangles]{Calverley11},
    \citet[diamonds]{BeckerBolton13}, \citet[squares]{DAloisio18},
    \citet[hexagons]{Davies18a} and \citet[stars]{Gallego21}. In all
    panels, different line-types refer to the predictions of our
    empirical modelling for different values of the $f_{\rm dc}$
    parameter (as indicated in the legend). The yellow shaded area
    represents the range of predictions for $f_{\rm dc}=0.1$ models
    with $\epsilon$ ranging from 0.05 to 0.15. Right and left panels
    refer to different choices of the limiting integration magnitude
    for the AGN-LF, as indicated in the upper label (see text for a
    more detailed discussion on our choices for $f_{\rm esc}$ and
    $M_{\rm lim}$).}\label{back_evo}
\end{figure*}

Our estimated evolution of the $z>5$ AGN-LF can be translated into a
prediction for the AGN contribution to the observed photoionization
rate and ionizing photons volume emissivity in the Early Universe
(Fig.~\ref{back_evo}), using the same formalism as described in
\citet{Cristiani16}. We summarise the main steps in the
following. Following \citet{HaardtMadau12}, we numerically solved the
equations of radiative transfer to get the photoionization rate
$\Gamma$:

\begin{equation}
\Gamma(z) = 4 \pi ~ \int_{\nu_H}^{\nu_{\rm up}} \frac{J(\nu,z)}{h_p \nu} \sigma_{HI}(\nu) d\nu
\end{equation}

\noindent
In the previous equation, $J(\nu,z)$ is the background intensity
computed as:

\begin{equation} \label{eq:Jnu}
J(\nu,z) = c/4 \pi \int_{z}^{\infty} \epsilon_{\nu_1}(z_1)
e^{-\tau_{\rm e}} \frac{(1+z)^3}{(1+z_1)^3} \, |\frac{dt}{dz_1}| \,
dz_1
\end{equation}

\noindent
where $\tau_{\rm e}(\nu,z,z_1)$ is the effective opacity between $z$
and $z_1$:

\begin{equation} \label{eq:teff}
\tau_{\rm e}(\nu,z,z_1) = \int_{z}^{z_1} dz_2 \int_{0}^\infty  dN_{HI} f(N_{HI},z_2) (1-e^{-\tau_{\rm c}(\nu_2)})
\end{equation}

\noindent
and $\tau_{\rm c}$ is the continuum optical depth. Moreover we also
estimate the comoving density of ionizing photons as:

\begin{equation}\label{eq:intlf}
\dot{N}_{\rm ion}(z) = \int_{\nu_H}^{\nu_{\rm up}}  \frac{\rho_\nu}{h_p \nu} d\nu
\end{equation}

\noindent
In all previous equations, $\nu_i = \nu \frac{1+z_i}{1+z}$, $\nu_H$ is
the frequency corresponding to $912$ {\AA} and $\nu_{\rm up} = 4
\nu_H$; $f(N_{HI},z)$ is the bivariate distribution of absorbers as in
\citet{BeckerBolton13}; $\epsilon_\nu(z)$ and $\rho_\nu$ represent the
proper and comoving volume emissivity (at frequency $\nu$),
respectively, that can be computed by integrating the AGN-LF
$\Phi(L,z)$, e.g.:

\begin{equation}\label{eq:lfint}
\rho_\nu = \int_{M_{\rm min}}^\infty f^{\rm AGN}_{\rm esc}(L,z) \, \Phi(L,z) \, L_\nu(L) \, dL
\end{equation}

\noindent
We assume a universal QSO/AGN broken power-law spectral shape of the
form $f_{\nu} \propto \nu^{-\gamma}$: in detail, we use $\gamma=-0.70$
in the wavelength range 500 {\AA} $< \lambda <$ 1000 {\AA}
\citep{Shull12, Lusso15} and $\gamma=-2$ at shorter wavelengths
\citep{Telfer02}. A key parameter is the integration depth, $M_{\rm
  min}$, which limits the number of sources that are included in the
computation. $f_{\rm esc}^{\rm AGN}$ represents the escape fraction
(i.e. the fraction of ionizing photons produced by the source that are
able to escape the galaxy and ionize the intergalactic medium). In
principle, the escape fraction could be a function of both the
luminosity of the object and its redshift (as well as of other
physical properties). In this paper we consider a fixed $f_{\rm
  esc}^{\rm AGN}=0.75$ value, based on the \citet{Cristiani16}
estimate for $M_{1450} \lesssim -27.5$ QSOs.

Fig.~\ref{back_evo} shows the predicted evolution of the
photoionization rate and ionizing photons volume emissivity for
different $f_{\rm dc}$ values, ranging from 5 to 35 percent. We
present two different scenarios, based on different assumptions for
the limiting magnitude. In the left panels, we show a conservative
scenario, where we consider only ionizing photons coming from QSOs
(i.e. $M_{\rm lim}=M_{1450}<-23$). On the right panel, instead, we
discuss the predictions on a more speculative scenario, where we
assume that our modelling hold up to $M_{\rm lim}<-18$ and that the
derived properties of bright AGNs are representative of objects living
on the faint-end of the LF as well. In particular, we consider the
same $f_{\rm esc}^{\rm AGN}=0.75$, derived for bright QSOs, over the
whole luminosity range, i.e. we imply that fainter AGNs resemble
scaled-down versions of the most powerful lighthouses in the
Universe. Low-z observations of AGNs around and fainter than the knee
of the LF suggest that these sources are not dramatically different
from bright counterparts \citep{Stevans14, Boutsia18, Grazian18}.

In general, large $f_{\rm dc}\gtrsim0.2$ values imply a fast evolution
of the AGN-LF and a fast drop of both the photon emissivity and
photoionization rate at $z>5$, well below the present available
constraints. Similarly, small $f_{\rm dc}<0.1$ correspond to a
negligible evolution of the AGN-LF and a flatter evolution for the
background. Therefore, intermediate values ($0.05<f_{\rm dc}<0.2$)
result in $\dot{N}_{\rm ion}$ and $\Gamma$ predictions that are the
most consistent with the observed evolutionary trends. Changing
$\epsilon$ within reasonable values (i.e. from 0.05 to 0.15) provides
predictions that are qualitatively consistent. The yellow shaded area
represents the span of models with fixed $f_{\rm dc}=0.1$ and a
variable $\epsilon$ between 0.05 and 0.15 (upper and lower envelope
respectively), and it is representative for other $f_{\rm dc}$
choices.

As already shown in our previous work \citep{Boutsia21}, the relative
normalization of predictions and data is tightly linked to the
integration depth assumed on the AGN-LF. Moreover, our modelling
neglects completely the contribution of star-forming galaxies at
comparable redshifts: those sources, although less efficient in
producing ionizing photons, have space densities much larger than the
AGN population and may supply a relevant contribution to the total
ionizing background \citep{Fontanot12a,Cristiani16}. Nonetheless, our
modelling for $f_{\rm dc}=0.1$ shows that, if sources fainter than
$M_{1450} = -23$ (and up to $M_{1450}<-18$) are taken into account,
Eddington-accreting AGNs at $z>5$ can provide enough ionizing photons
to account for the observed background. Such a deep integration limit,
tied with the assumption that the high $f_{\rm esc}^{\rm AGN}$
observed in the brightest QSO does not dramatically drop for faint
AGNs is crucial in our framework. As an alternative interpretation,
$M_{\rm lim}$ can be viewed as the limiting magnitude of the AGN
population characterized by a $f_{\rm esc}^{\rm AGN}$ comparable to
bright QSOs, that is to say that a luminosity dependent $f_{\rm
  esc}^{\rm AGN}$ prescription would naturally predict an equivalent
$M_{\rm lim}$. However, as clearly shown in Fig.~\ref{back_evo}, both
a deep integration of the AGN-LF and a high $f_{\rm esc}^{\rm AGN}$
are required for the AGN population to provide a ionizing photon space
density comparable with the estimated background at $z\simeq5$, which
represents the starting point of our analysis.

\subsection{Reionization}
\subsubsection{Modelling of evolution of the ionized fraction}
Additional insight on the photon budget in the EoR is provided by the
redshift evolution of the neutral fraction of the dominant baryonic
components of the Universe, Helium and Hydrogen ($x_{\rm HeIII}$ and
$x_{\rm HII}$, respectively). The evolution of the neutral fraction is
linked\footnote{We also assume that single ionized Helium evolves as
  ionized Hydrogen.} to the corresponding filling factors ($Q_{\rm
  HeIII}$ and $Q_{\rm HII}$); we model the $Q$ evolution following two
different approaches. The first one represents the standard approach
in the literature: following \citet{MHR99}, we assume that
reionization is an {\it homogeneous} process and that $Q$s obey the
equation describing the evolution of the filling factors:

\begin{equation}\label{eq:q2}
\dot{Q} = \frac{\dot{N}_{\rm ion}}{\langle n \rangle} - \frac{Q}{\langle t_{\rm rec} \rangle}
\end{equation}

\noindent
where $\dot{N}_{\rm ion}$ is the comoving density of ionizing photons
for each species (i.e. between 1 and 4 Rd for Hydrogen, between 4 and
16 Rd for Helium); $\langle n \rangle$ is the mean comoving density of
atoms of the considered species (with $\langle n_{\rm He} \rangle =
\langle n_H \rangle /12.$) and $\langle t_{\rm rec} \rangle$ is the
volume-averaged recombination rate of the species \citep[see
  e.g.][]{MadauHaardt15}:

\begin{displaymath}
\langle t^{\rm HII}_{\rm rec} \rangle^{-1} =  C(z)  \, (1 + \chi) \, \langle n_H \rangle  \,  \alpha_{\rm HII}(T) \, (1+z)^3 
\end{displaymath}

\begin{displaymath}
\langle t^{\rm HeIII}_{\rm rec} \rangle^{-1} =  C(z)  \,  Z \, (1 + 2\chi) \, \langle n_H \rangle \, \alpha_{\rm HeIII}(T/Z^2) \, (1+z)^3 
\end{displaymath}

\noindent
where $\chi = Y/[4(1-Y)]$ includes photoelectrons from He$_{\rm II}$;
$Y$ is the primordial Helium mass fraction, $Z=2$ is the ionic charge;
$C(z)= 2.9 [ (1+z)/6 ]^{-1.1}$ is the redshift dependent clumping
factor (that we assume being the same for both species) as in
\citet{MadauHaardt15} and $\alpha$s represent the case B recombination
coefficient for H$_{\rm II}$ and He$_{\rm III}$ \citep{HuiGnedin97}:

\begin{displaymath}
\langle \alpha_{\rm HII} \rangle = 2.753 \times 10^{-14} \frac{(315614/T)^{1.5}}{ \left[ 1 + (115188/T)^{0.407} \right]^{-2.242} }
\end{displaymath}

\begin{displaymath}
\langle \alpha_{\rm HeIII} \rangle = 5.506 \times 10^{-14}  \frac{(1263030/T)^{1.5}}{ \left[ 1 + (460960/T)^{0.407} \right]^{-2.242} }
\end{displaymath}

\noindent
In practice, for the purpose of this work, we assume a fixed value for
the temperature $T=10^4 $K, which is appropriate for ionizing regions
around QSOs.
\begin{figure}
  \centerline{ \includegraphics[width=9cm]{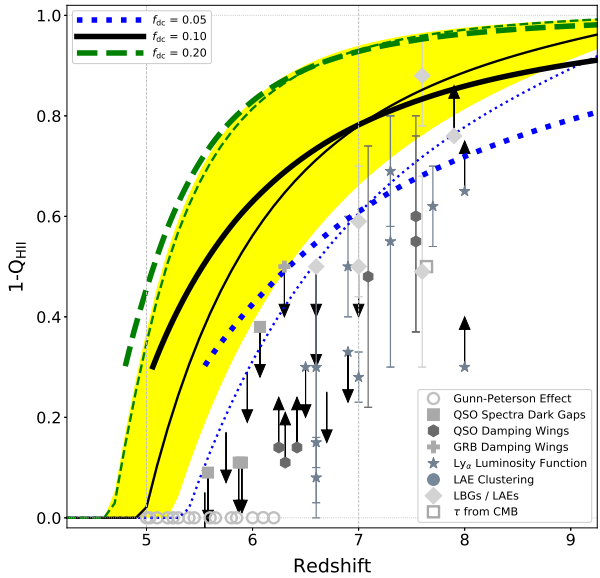} }
  \caption{Redshift evolution of the neutral hydrogen
    fraction. Different line-types and colours refer to the
    predictions of our empirical modelling for different values of the
    $f_{\rm dc}$ parameter (as indicated in the legend). Thin and
    thick line refer to the predictions of the {\it homogeneous} and
    {\it bubble} models, respectively (see main text for more
    details). The yellow shaded area represents the range of
    predictions for $f_{\rm dc}=0.1$ models with $\epsilon$ ranging
    from 0.05 to 0.15. Grey symbols show the available observational
    constraints listed in Table~\ref{tab:xhi}.}\label{fig:q2}
\end{figure}
\begin{figure}
  \centerline{ \includegraphics[width=9cm]{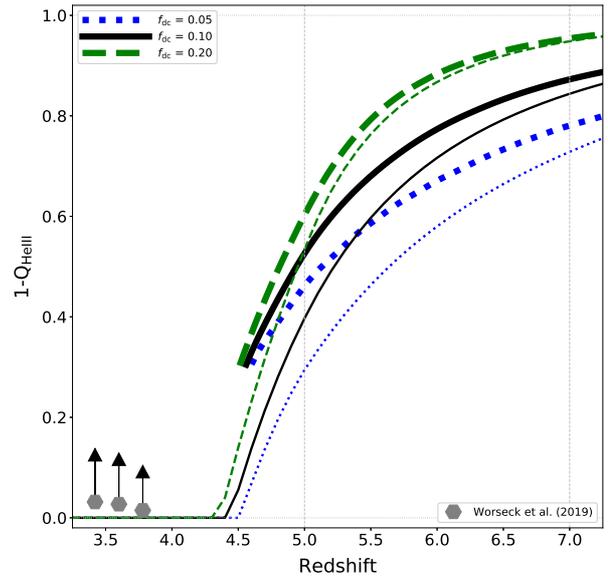} }
  \caption{Redshift evolution of the double ionized Helium fraction
    ($Q_{\rm HeIII}$). Different line-types and colours refer to the
    predictions of our empirical modelling for different values of the
    $f_{\rm dc}$ parameter (as indicated in the legend). Thin lines
    refer to a model assuming {\it homogeneous} Helium reionization,
    while thick lines show the predictions for the {\it bubble} model
    for the evolution of the ionization front of each luminous AGN/QSO
    (see text for more details). Grey symbols mark the observational
    determinations by \citet{Worseck19}.}\label{fig:qHe}
\end{figure}

The {\it homogeneous} model might not be able to recover some of the
details of the reionization process. In particular, the extent of the
EoR depends on its topology, i.e. on the spatial distribution of the
ionizing sources and their clustering. This is especially important
for our hypothesis of an AGN-driven reionization, as AGNs are more
sparse and rare sources, with respect, e.g., to star-forming galaxies
at comparable redshifts. In order to take these effects into account,
while keeping our modelling simple, we develop what we call the ``{\it
  bubble}'' model. We assume that each AGN develops an (almost)
spherical ionized region, and we thus model the evolution of the
associated ionization front as a function of its luminosity. Following
\citet{Khrykin16}, we assume that an AGN of luminosity $M_{1450}$ is
able to affect a spherical volume of radius:

\begin{equation}\label{eq:kry}
  R_{\rm IF}(M_{1450}) = R_S(M_{1450}) \left [ 1- \exp \left( - \frac{t_{\rm lf}}{x_Q t_{\rm rec}} \right)\right]^{1/3}
\end{equation}

\noindent
where $R_S$ represents the classical Str\"omgren radius:

\begin{equation}
R_S(M_{1450}) = \frac{3}{4 \pi} \frac{\dot{N}_{\rm ion} (M_{1450})}{\langle n \rangle / t_{\rm rec}}
\end{equation}

\noindent
In the following, we assume that each new QSO episode carves a ionized
bubble from a non-ionized medium (i.e. $x_Q=1$, which correspond
either to a full neutral hydrogen or to a single ionized helium). For
the sake of simplicity, we consider a fixed QSO lifetime $t_{\rm lf}$,
defined as the $f_{\rm dc}$ fraction of the time interval between
$z=5$ and $z=7$. This choice implies that, for $f_{\rm dc} = 0.1$,
$t_{\rm lf}$ roughly corresponds to a Salpeter time (i.e. 45 Myrs).

At each redshift, we then compute the size of ionized bubbles as a
function of $M_{1450}$; by combining the corresponding spherical
volumes with the expected space density of sources at that given
luminosity, we estimate the volume fraction of the newly ionized
medium. We thus change the source term in Eq.~\ref{eq:q2} with this
estimate to test for changes in the $Q$ evolution.

It is worth stressing that the {\it bubble} model most likely break
down for values of $Q$ approaching unity. Indeed, in our calculations
we are implicitly assuming that each new HII (HeIII) bubble starts in
an homogeneous neutral (HeII) medium, with no interaction with nearby
similar structures. However, for large $Q$ values this is no longer
the case, as both physical mechanisms (like bubble percolation) and
geometrical considerations (like in the case of AGN clustering, which
allows for a QSOs shining in a medium that has been already partially
ionized) start to be relevant. As an example, \citet{DoussotSemelin22}
study the effect of percolation on statistics of ionized bubble size
distribution and find that percolation has a relevant effect for
$Q>0.7$. It is not easy to assess the effect of these mechanisms on
the {\it bubble} model. On one hand, both percolation and clustering
favour the formation of larger ionization fronts, that should speed up
the reionization process (i.e a sharper rise of $Q$ to
unity). Nonetheless, if AGN sources are highly clustered, this implies
that could exist neutral regions far enough from the closest AGN to be
able to survive up to low redshifts. The relative contribution of
these two different scenarios is impossible to determine in our
simplified approach, so that we prefer to limit the {\it bubble} model
to $Q<0.7$ in the following analysis.

\subsubsection{Hydrogen Reionization}
In Fig.~\ref{fig:q2}, we compare the available constraints on $x_{\rm
  HI}$, coming from different techniques (see Table~\ref{tab:xhi} for
detailed references), with the prediction from our {\it homogeneous}
model (thin lines) for different $f_{\rm dc}$ choices.
Fig.~\ref{fig:q2} clearly shows that in our reference frame only small
$f_{\rm dc}$ values are compatible with the data, while larger values
require some additional sources of ionizing photons (i.e. star forming
galaxies) to close the photon budget. Alternatively, $\epsilon>0.1$
values are required for models with intermediate $f_{\rm dc} \simeq 0.1$
(lower boundary of the yellow region). Wen considering the {\it
  bubble} model, the overall predictions do not change considerably;
the largest differences are seen for $f_{\rm dc}<0.2$, with the {\it
  bubble} model predicting slightly more extended EoRs and
lower reionization redshifts.

A comparison of Fig.~\ref{back_evo} and~\ref{fig:q2} suggests that,
in order to reproduce the $x_{\rm HI}$ evolution, a rather flat and
constant $\dot{N}_{\rm ion}$ background is needed \citep[at the level
  of the observed value at $z\simeq5$, see also][]{Madau17}. In the
context of our model that considers only the AGN contribution, such a
background can be achieved only with a slowly evolving AGN LF
(i.e. $f_{\rm dc}<0.05$ values). Nonetheless, a flat ionizing
background seems to be in tension with the available constraints on
the photoionization rate at $z\simeq6$. Such an apparent tension is
mainly due to the fact that Eq.~\ref{eq:teff} holds only for a fully
ionized inter-galactic medium. Indeed, \citet{Puchwein19} show that a
more detailed treatment of the effective opacity during the EoR,
taking into account the inhomogeneity of the medium (i.e. the presence
of regions of neutral hydrogen and helium), leads to a rapid evolution
of the mean free path of ionizing photons \citep{Becker21}. The
improved modelling leads naturally to a sharp discontinuity in
$\Gamma$ during the EoR, which fits nicely the highest redshift
observational determination also in the case of an almost flat
background (their Fig.~3).

\subsubsection{Helium reionization}
A critical test on models for AGN-driven Hydrogen reionization comes
from the additional constraints based on the reionization history of
the Helium component \citep[see e.g.][]{FurlanettoOh08}. Helium
reionization is believed to be completed by $z\gtrsim3$, and sustained
by photons above 4 Ry, provided by the growing population of AGNs at
$z<4$ \citep{WyitheLoeb03}. Significant fluctuations of the He$_{\rm
  III}$ effective optical depths have been detected at $3<z<4$
\citep[e.g.][]{Worseck16, Worseck19}, which cannot be explained by
models assuming an uniform mean free path for ionizing photons
\citep{FurlanettoDixon10} and suggest incomplete Helium reionization
at these redshifts \citep{Worseck11}. Indeed, our {\it homogeneous}
model predicts relatively early Helium reionization redshifts ($z>4$ -
Fig.~\ref{fig:qHe}, thin lines), in agreement with,
e.g.,\citet{MadauHaardt15}. On the other hand, the {\it bubble} model
provides a quite different $Q$ evolution (Fig.~\ref{fig:qHe} - thick
lines): with respect to the {\it homogeneous} model, for all $f_{\rm
  dc}$ values, the Helium reionization is more extended and
systematically delayed to lower redshifts. These trends decrease for
larger $f_{\rm dc}$ due to the fact that they corresponds to larger
bubble sizes at fixed luminosity (which thus deviate less from the
homogeneous approximation), and the increase in size compensate the
smaller space densities associated with the faster LF evolution
(Sec.~\ref{sec:lf}). It is worth comparing these predictions with
recent estimates of QSO lifetimes by \citet{Khrykin21}. They apply
Eq.~\ref{eq:kry} to the analysis of Helium proximity zones in
absorption spectra of individual QSOs at $3<z<4$ QSOs, and find that
the spectroscopic data are mostly consistent with short lifetimes in
the range of $t_{\rm lf} \sim$ 0.5-20 Myr (with a mean value of 1.65
Myr). In the redshift range $5<z<7$, lifetimes $\lesssim$ 20 Myr
(i.e. half of a Salpeter time) correspond to $f_{\rm dc} \lesssim
0.05$ implying an extended Helium reionization, starting at $z>5$ and
being completed at $z<3$.

The most striking difference between the {\it homogeneous} and the
{\it bubble} model for Helium reionization is the systematic drift of
$Q$ evolution to lower redshifts at all cosmic epochs, while in the
case of Hydrogen reionization the two models provide more similar $Q$
evolutions. It is not easy to understand the origin of this effect, as
it is due to a combination of the smaller number of ionizing photons
available, the smaller number of HeII atoms and the shorter
recombination timescales. We check that our results are robust against
reasonable changes in the recombination timescale and clumping factors,
which suggests that the smaller comoving density of ionizing photons
for Helium with respect to Hydrogen plays the larger role.

\subsection{Evolution of high-z BH seeds}
Our framework also allows us to explore the predicted distribution of
BH masses at $z\gtrsim5$ and to check under which conditions it
complies with the most recent theoretical expectations for the
properties of primordial BH seeds. The Initial Mass Function of BHs
(BH-IMF) is expected to have a peak at stellar-like masses
(i.e. 10-100 $M_\odot$ - light seed BHs) corresponding to the remnants
of stellar evolution, including the elusive Population III stars
\citep[see e.g.][]{GarciaBellido21, Sureda21}. It is important to keep
in mind that, in order to reach a $\sim10^9 M_\odot$ SMBH at
$z\simeq7$, a $\sim$ 100 $M_\odot$ Pop III seed BH would need 0.8 Gyrs
of Eddington accretion, i.e. $f_{\rm dc}\simeq$1. Nonetheless, the
BH-IMF most likely spans the full range of 10-10$^6 M_\odot$: massive
seed BHs may result from the near-isothermal collapse of a chemically
pristine massive gas cloud \citep[see e.g.][]{Ferrara14}, from stellar
mergers in ultradense star clusters \citep{DevecchiVolonteri09}, or
even being the relic of a primordial BH population \citep[see
  e.g.][]{YuTremaine02}. These very massive seeds may only form in
highly biased regions of the Universe, moreover, only a small minority
of them are bound to evolve into SMBHs, the others being expected to
remain lower-mass BHs at the centre of dwarf satellites
\citep{Valiante16}. Therefore, even these models with such massive
seeds seem to require high $f_{\rm dc}$ values to explain the observed
$z\simeq7$ SMBHs \citep{TanakaHaiman09}. Indeed, the situation is
further complicated by the possibility of hyper-Eddington accretion
\citep{Inayoshi16, Takeo18}: as a result of a short ($0.2<f_{\rm
  dc}<0.3$) phase of the order of 500 $\dot{M}_{\rm edd}$ most of the
stellar mass remnants would reach masses $\sim$ 2 $\times 10^5
M_\odot$, independently of their initial seed mass, thus providing a
larger sample of massive SMBHs available for powering luminous sources
at high-z \citep{Inayoshi20}.

\begin{figure}
  \centerline{ \includegraphics[width=9cm]{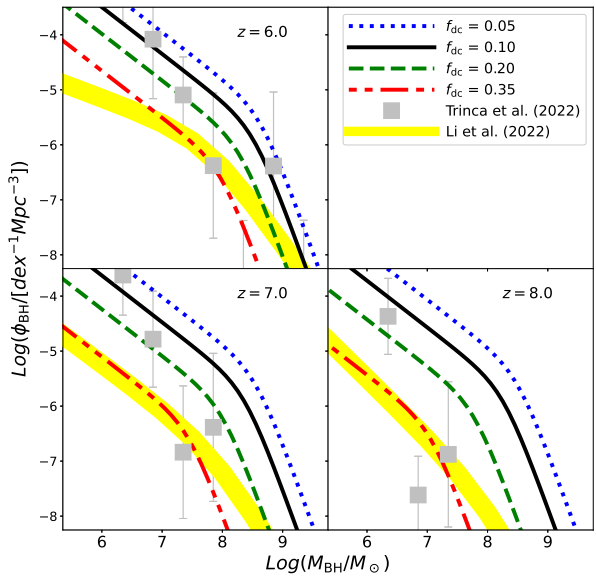} }
  \caption{Total (active+inactive) BHMF at $z>5$. Different line-types
    refer to the predictions of our empirical modelling for different
    values of the $f_{\rm dc}$ parameter (as indicated in the
    legend). The grey squares refer to the theoretical predictions for
    the evolution of direct collapse BHs from \citet{Trinca22}, while
    the yellow area represents the BHMFs predicted by the model
    proposed by \citet{Li22}. }\label{ratio}
\end{figure}

Additional constraints on $f_{\rm dc}$ can be thus derived from
considering the predicted evolution for the total BHMF. We derived
this quantity from the space density of active SMBHs multiplied by the
adequate $f_{\rm dc}$. As a benchmark, we compare our models
(Fig.~\ref{ratio}) against the BH space densities predicted by the
semi-analytic codes like {\sc cat} \citep[Cosmic Archaeology
  Tool][grey squares]{Trinca22} and the model proposed by
\citet{Li22}. The main differences among the two approaches lies in
the assumed starting distribution of light and heavy seeds and in the
allowed range of Eddington ratios (most notably \citealt{Li22} require
several episodes of Super-Eddington accretion, while the best-fit {\sc
  cat} model is Eddington-limited). Despite their differences the two
codes provide a consistent prediction for the space density of the
more massive BHs in a wide redshift range, while they start to diverge
at the low-mass end of the BHMF, with {\sc cat} predicting
sistematically higher space densities.

While the lower space densities at the faint end predicted by the
\citet{Li22} model favour\footnote{However, we notice that including
episodes of super-Eddington accretion has the effect of speeding up
the evolution of the BHMF, with respect to our modelling.}  large
$f_{\rm dc} \simeq 0.35$ values, our models with $0.1<f_{\rm dc}<0.2$
are mostly consistent with {\sc cat} over the $6<z<10$ redshift
range. Models with larger $f_{\rm dc}$ values start to be in tension
with {\sc cat} already at $z=6$, while models with smaller $f_{\rm
  dc}$ values tend to overpredict the space density of SMBHs. While
interpreting these plots it is also important to keep in mind that we
define $f_{\rm dc}$ based on a reference redshift range (that is
$5<z<7$); moving outside this range to higher redshifts implies that
$f_{\rm dc}$ should be seen as a fixed time interval, more than a time
fraction (and in particular $5<z<7$ roughly corresponds to a cosmic
time interval ten times the Salpeter time).

\section{Conclusions}\label{sec:concl}

We develop an empirical model aimed at studying the redshift evolution
of the AGN/QSO-LF under the hypothesis that all SMBHs at $z\gtrsim5$
(where we fixed our starting space densities using available
estimates) accrete at Eddington rate, which allows us to give redshift
estimates also for the BHMF (both for the active and total
population). Such a prediction is a fundamental constraint for our
approach, since the existence of very massive SMBHs at such high
redshift is challenging for current models of early BH
``seeding''. The existence of a few $10^9 M_\odot$ SMBHs, powering the
observed high-luminosity sources at $z\simeq7$ \citep{Mortlock11,
  Banados18, Matsuoka19, Wang21}, can be generally reconciled with the
expected formation of massive structures at high redshift, by assuming
that these are the end product of the rare seeds experiencing very
efficient accretion \citep[see e.g.][]{DiMatteo12}. However, if SMBHs
of comparable mass are common also at higher redshifts, their
accretion histories are increasingly difficult to reconstruct.

We can roughly divide the predictions of this model into three
separate regimes. $f_{\rm dc} \gtrsim 0.2$ scenarios correspond to
fast evolving AGN-LF and negligible space densities of $M_{\rm SMBH}
\gtrsim 10^8 M_\odot$ objects at $z>6$. These models are consistent
with conservative estimates for the high-z growth of BH seeds
\citep{TanakaHaiman09}, but they under-predict the space densities
implied by more sophisticated models of seed growth
\citep{Trinca22, Li22}. Overall, they also predict a vanishing contribution
of the AGN population to the observed ionizing background
at $z>5$, independently of the integration magnitude $M_{\rm
  lim}$.

The evolution of the AGN LF for intermediate duty cycles ($f_{\rm dc}
\simeq 0.1$) implies BHMFs in good agreement with the expectations of
the {\sc cat} semi-analytic model for the growth of BH seeds in the
early Universe. Moreover, they are also able to provide a relevant
contribution to the observed ionizing background and photoionization
rate, in good agreement with the data, if a deep integration limit
($M_{\rm lim}$) is assumed. Nonetheless, they also predict an
important redshift evolution of the AGN space density/ionizing
background, which is inconsistent with the observed evolution of the
neutral fraction (thus suggesting the need for an additional source of
ionizing photons in the EoR).

Finally, $f_{\rm dc} \lesssim 0.05$ scenarios feature a slow evolution of
the AGN-LF evolution, which correspond to a rather shallow evolution
of the ionizing background at high-z. In these models the AGN
population alone provides enough ionizing photons to account for the
evolution of the Hydrogen neutral fraction. Moreover, such low $f_{\rm
  dc}$ values imply QSO lifetimes in good agreement with estimates at
lower redshift \citep{Khrykin21}. However, under the hypothesis of
homogeneous reionization, such models provide too high redshifts for
Helium reionization ($z>4$). We show that a simple model taking
into account the topology of reionization (e.g. by following the
growth of ionized bubbles around accreting SMBHs) predicts more
extended Helium reionization histories, thus easing the tension
between AGN-driven Hydrogen reionization scenarios and available data
on Helium reionization.

However, the slow evolution of the LF translates into space densities
for $M_{\rm SMBH} \sim 10^{8-9} M_\odot$ sources relatively high even
at $z>6$. Such space densities would be in tension with most models of
SMBHs seeding and accretion using Eddington accretion. Nonetheless,
these results could be reconciled with theoretical expectations by
assuming a relatively short period of super/hyper-Eddington accretion
\citep[see also][]{Pezzulli16} at $z>7$ that could easily bring a good
fraction of the light seed into the $\sim$ 10$^5 M_\odot$ mass range,
thus changing the shape of the highest-z BH-IMF. If such a scenario
holds, we may have enough massive SMBH at $5<z<7$ to account for
these relatively low duty cycles.

It is extremely difficult to disentangle these scenarios based on the
available data, which become sparse at $z \gtrsim 5$. Our modelling
provides a reference frame to investigate the role of AGN/QSO in the
EoR, based on a number of assumptions, the most relevant being
the idea that the properties of faint sources (most notably their
$f_{\rm esc}^{\rm AGN}$) can be derived from their bright QSO
counterparts. Starting from this assumption, we could place
interesting constraints to the contribution of the AGN population to
the photon budget during the EoR: depending on the assumed $M_{\rm
  lim}$ the AGNs/QSOs move from being a relevant contributor into
being the dominant population. This in turn implies that, while
waiting for the James Webb Space Telescope to provide unprecedented
constraints on the evolution of the high-z AGN-LF, coordinated efforts
like QUBRICS represent excellent pathfinders for our understanding of
the processes at play in such early epochs.

\section*{Acknowledgements}
We warmly acknowledge J.~Sureda and A.~Trinca for sharing the
predictions of their theoretical models and for useful discussions on
the rise of primordial BH seeds.  A.G. and F.F. acknowledge support
from PRIN MIUR project ``Black Hole winds and the Baryon Life Cycle of
Galaxies: the stone-guest at the galaxy evolution supper'', contract
2017-PH3WAT.

\section*{Data Availability}
Predictions for the $z>5$ AGN-LF evolution and/or relative AGN
ionizing background for a variety of parameter combinations will be
shared on reasonable request to the corresponding author.

\bibliographystyle{mnras} \bibliography{fontanot}

\appendix

\section{Neutral fraction estimates}

Table~\ref{tab:xhi} collect available estimates for the evolution of
the neutral fraction at $z>5$ coming from different techniques (as
listed in the title of the different sections).

\begin{table*}
  \centering
  \caption{Constraints on $x_{HI}$ from the literature (updated from \citealt{Bouwens15a})}
    \label{tab:xhi}
    \renewcommand{\footnoterule}{}
    \begin{tabular}{ccc}
      \hline
      {\bf Redshift}   & $x_{HI}$     & {\bf Reference} \\
      \hline
      \multicolumn{3}{c}{{\bf Gunn-Peterson Effect} (data are in units of $ 10^{-5} $)} \\
       & & \\
      5.03                & $5.49^{+1.42}_{-1.65}$   & \citet{Fan06} \\
      5.25                & $6.70^{+2.07}_{-2.44}$    & " \\
      5.45                & $6.77^{+2.47}_{-3.01}$    & " \\
      5.65                & $8.60^{+3.65}_{-4.60}$    & " \\
      5.85                & $12.00^{+4.08}_{-4.90}$   & " \\ 
      6.10                & $43.^{+30.}_{-30.}$       & " \\ 
      5.40                & $5.71^{+0.59}_{-1.21}$    & \citet{Yang20} \\
      5.60                & $7.61^{+1.61}_{-0.75}$    & " \\
      5.80                & $8.8^{+1.8}_{-1.2}$       & " \\
      6.00                & $11.4^{+5.5}_{-1.9}$      & " \\
      6.20                & $10.3^{+5.5}_{-1.1}$      & " \\      
      5                   & $3.020^{+0.230}_{-0.058}$ & \citet{Bosman21} \\ 
      5.1                 & $3.336^{+0.064}_{-0.164}$ & " \\ 
      5.2                 & $3.636^{+0.131}_{-0.095}$ & " \\ 
      5.3                 & $3.598^{+0.556}_{-0.145}$ & " \\ 
      \hline
      \multicolumn{3}{c}{\bf Dark Pixel Fraction in Quasar Spectra} \\
       & & \\
      5.58                & $<0.09$                    & \citet{McGreer15}\\
      5.87                & $<0.11$                    & " \\
      6.07                & $<0.38^{+0.20}$             & " \\
      5.9                 & $\lesssim0.11$             & \citet{GreigMesinger17} \\
      5.55                & $<0.05$                    & \citet{Zhu22} \\
      5.75                & $<0.17$                    & " \\
      5.95                & $<0.29$                    & " \\
      \hline
       \multicolumn{3}{c}{\bf Ly$\alpha$ Damping Wing (QSOs)} \\
       & & \\
      6.247               & $\gtrsim0.14$         & \citet{Schroeder13} \\
      6.308               & $\gtrsim0.11$         & " \\
      6.419               & $\gtrsim0.14$         & " \\
      7.09                & $0.48^{+0.26}_{-0.26}$  & \citet{Davies18}\\
      7.54                & $0.60^{+0.20}_{-0.23}$  & " \\
      7.54                & $0.55^{+0.21}_{-0.18}$  & \citet{Banados18}\\
      7.29                & $0.49^{+0.11}_{-0.11}$  & \citet{Greig22} \\
      \hline
   \end{tabular}
    \begin{tabular}{ccc}
      \hline
      {\bf Redshift}   & $x_{HI}$     & {\bf Reference} \\
      \hline
      \multicolumn{3}{c}{\bf Ly$\alpha$ Damping Wing (GRBs)} \\
       & & \\
      6.3                 & $\leq0.5$             & \citet{Totani06} \\
      \hline
      \multicolumn{3}{c}{\bf Ly$\alpha$ Luminosity Function} \\
       & & \\
      6.5                 & $\lesssim0.3$         & \citet{Malhotra04}\\ 
      6.6                 & $0.08^{+0.08}_{-0.05}$  & \citet{Morales21} \\
      7.0                 & $0.28^{+0.05}_{-0.05}$  & " \\
      7.3                 & $0.69^{+0.11}_{-0.11}$  & " \\
      6.6                 & $0.3^{+0.2}_{-0.2}$     & \citet{Konno18} \\
      6.6                 & $0.15^{+0.15}_{-0.15}$  & \citet{Ouchi18} \\
      6.9                 & $0.4-0.6$             & \citet{Zheng17} \\
      6.9                 & $<0.33$               & \citet{Wold22} \\
      7.3                 & $0.55^{+0.25}_{-0.25}$  & \citet{Konno14} \\
      7.7                 & $0.62^{+0.08}_{-0.08}$  & \citet{Faisst14} \\
      $\sim8$             & $\gtrsim0.3$          & \citet{Tilvi14} \\
      8                   & $>0.65$               & \citet{Schenker14} \\
      6.6                 & $0.3^{+0.1}_{-0.1}$     & \citet{Ning22} \\
      \hline
      \multicolumn{3}{c}{\bf Ly$\alpha$ Emitting Galaxies / Lyman Break Galaxies} \\
       & & \\
      $\sim7$             & $0.59^{+0.11}_{-0.15}$  & \citet{Mason18} \\
      $7.6 \pm 0.6$       & $0.88^{+0.08}_{-0.10}$  & \citet{Hoag19} \\
      7.6                 & $0.49^{+0.19}_{-0.19}$  & \citet{Jung20}\\
      $7.9 \pm 0.6$       & $>0.76$               & \citet{Mason19} \\
      6.6                 & $<0.4$                & \citet{Yoshioka22} \\
      $6.7 \pm 0.2$       & $<0.25$               & \citet{Bolan21} \\
      $7.6 \pm 0.6$       & $0.83^{+0.08}_{-0.11}$   & " \\      
      \hline
      \multicolumn{3}{c}{\bf Clustering of Ly$\alpha$ Emitting Galaxies} \\
       & & \\
      6.6                 & $<0.5$                & \citet{SobacchiMesinger15} \\          
      7.0                 & $\lesssim0.5$         & " \\
      \hline
      \multicolumn{3}{c}{\bf $\tau_{es}$ from CMB} \\
       & & \\
      $7.64\pm 0.74$      & $0.5$                 & \citet{Planck_cosmpar18} \\
      9.75                & $0.76^{+0.22}_{-0.27}$  & \citet{Dai19} \\
      \hline
    \end{tabular}
\end{table*}

\end{document}